\documentclass[lettersize,journal]{IEEEtran}
\usepackage{amsmath,amsfonts}
\usepackage{algorithmic}
\usepackage{array}
\usepackage[caption=false,font=normalsize,labelfont=sf,textfont=sf]{subfig}
\usepackage{textcomp}
\usepackage{stfloats}
\usepackage{url}
\usepackage{verbatim}
\usepackage{graphicx}
\def\BibTeX{{\rm B\kern-.05em{\sc i\kern-.025em b}\kern-.08em
    T\kern-.1667em\lower.7ex\hbox{E}\kern-.125emX}}
\usepackage{balance}
\usepackage{siunitx}

\begin{document}
\title{Hollow-core fiber for single-mode, low loss transmission of broadband UV light}
\author{D. Dorer, M. H. Frosz, S. Haze, M. Dei{\ss}, W. Schoch and J. Hecker Denschlag
\thanks{D. Dorer, S. Haze, M. Dei{\ss}, W. Schoch and J. Hecker Denschlag are with the Institute of Quantum Matter and the Center for Integrated Quantum Science and Technology IQ\textsuperscript{ST}, Ulm University, Albert-Einstein-Allee 45, 89081 Ulm, Germany \\}
\thanks{M. H. Frosz is with the Max Planck Institute for the Science of Light, Staudtstr. 2, 91058 Erlangen, Germany.}
}

\maketitle

\begin{abstract}
We report on an anti-resonant hollow-core fiber (AR-HCF) designed for stable transmission of laser light in a broad wavelength range of 250 nm to 450 nm. We tested for wavelengths of 300 nm and 320 nm. The characterized fiber shows a low transmission power attenuation of 0.13 dB/m and an excellent single-mode profile. The fiber maintains stable transmission after an exposure of tens of hours with up to 60 mW CW-laser light and shows no indication of solarization effects. We further tested its performance under bending and observed a small critical bending radius of about 6 cm. These characteristics make the presented fiber a
useful tool for many applications, especially in quantum optics labs where it may be instrumental
to improve on stability and compactness.
\end{abstract}


\section{Introduction}
\begin{figure*}[t!]
\centering
\includegraphics[width=5in]{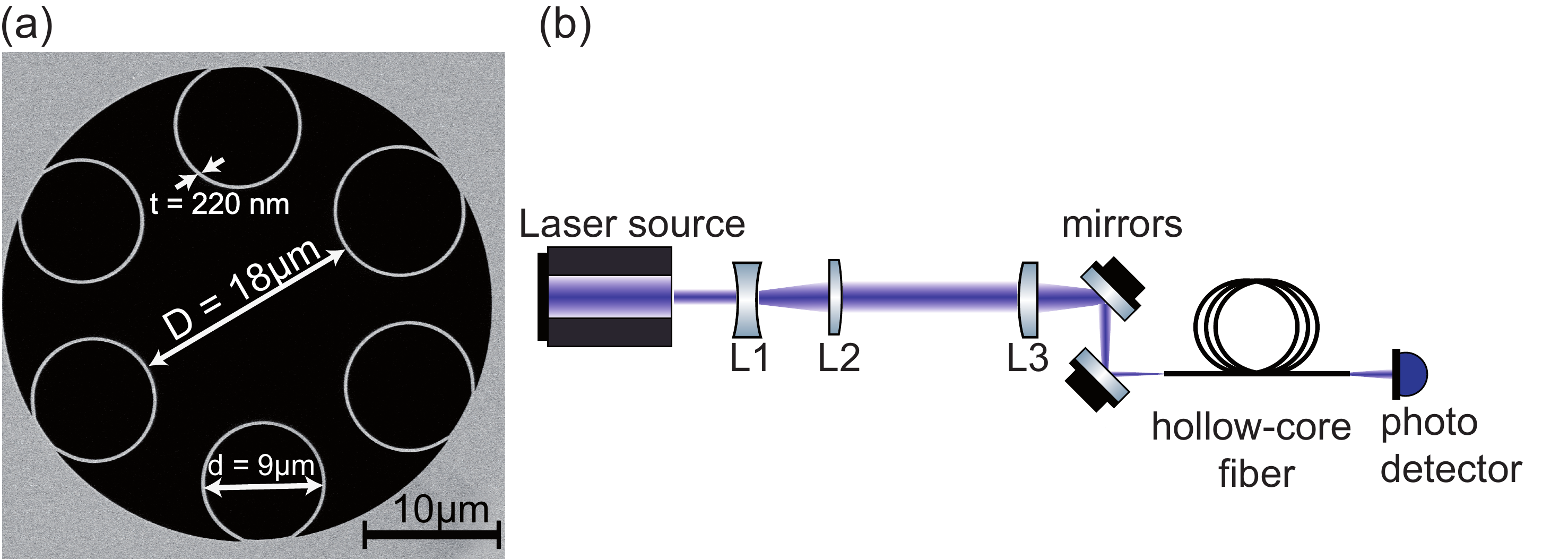}
\caption{(a) Scanning electron microscopy image of the fiber cross-section. The structure consists of a hollow-core diameter \emph{D} $\sim$ 18 µm diameter and six hollow capillaries with a diameter \emph{d} $\sim$ 9 µm each. The core wall thickness t is $\sim$ 220 nm. (b) Setup for fiber-coupling. The output beam from the laser source is expanded by lenses $L_1$ and $L_2$ to a waist of 1.9 mm and then subsequently focused down with lens $L_3$ with a focal length of 125 mm. With two highly-reflective mirrors, the beam is coupled into the fiber. The transmitted light is measured by a photodetector and converted into a laserpower signal. Alternatively, a camera is used for beam shape analysis.}
\label{Fig1:REM}
\end{figure*}

\IEEEPARstart{O}{ptical} fibers are well-established tools for stable transmission of laser light. For the visible to infrared wavelength range, silica-based optical fibers with a solid-core are routinely used for low-loss delivery of single-transverse-mode beams of high quality \cite{Tamura:18}. However, the applicability of solid-core fibers is severely compromised in the ultraviolet spectral range as they suffer irreversible degradation after even short periods of exposure to UV light \cite{Yamamoto:09}. This behavior is caused by UV-induced damage in the silica material - often referred to as solarization \cite{Skuja:01,Lancry:13, Dragomir:02}. On the other hand, for example, in atomic, molecular and optical physics, there is a vastly growing number of applications involving UV light, and particularly at a wavelength around 300 nm. As an example, Rydberg states of Cs, Rb, and K atoms (see, e.g. \cite{Deiss:19}) can be excited with such wavelengths in a single-photon excitation from the ground state. As another example, Doppler cooling and state readout of the trapped ions Be\textsuperscript{+}, Mg\textsuperscript{+}, and Yb\textsuperscript{+} works at the wavelengths 313 nm, 280 nm and 369 nm, respectively. Such platforms have a large potential for applications regarding, e.g., quantum computing, quantum simulation and metrology (see, e.g. \cite{Timoney2011, Ospelkaus2011, Leibfried2003,Zhang2020, Scholl2021, Bluvstein2022, Wolf2016}). Therefore, a good solution for stable and efficient, single-mode transmission of UV laser light could be very useful for such platforms and their applications. \\
One approach to overcome the solarization problem relies on hydrogen passivation \cite{Colombe:14,Marciniak2017}. In another approach, hollow-core fibers are used to transmit light having negligible overlap with the silica glass. The structures of such hollow-core fibers can be tailored to achieve single-mode delivery of UV light. An example of such a structure is the Kagom\'{e} type \cite{Gebert2014}. Recently, single-ring structures formed by capillaries surrounding the hollow-core have gained increasing interest because of both low-loss transmission and high-quality single-mode transmission achieved by modal filtering when the fiber structure is suitably designed \cite{Uebel:16, Gao:18}. An approach which targets a wavelength range around 300 nm and demonstrates low attenuation is reported in \cite{Gao:18} where a single-ring structure with six capillaries and a ratio between core diameter \emph{D} and capillary diameter \emph{d} of \emph{d/D} = 0.51 is used. We employ a similar design, however, our wall thickness of the ring capillaries \emph{t} $\sim$ 220 nm is about three times smaller, see Fig. \ref{Fig1:REM}(a). The wall thickness of the ring capillaries, which we call  in the following core-wall thickness, is crucial because it determines the wavelength of the loss resonances according to $\lambda_{q} = 2t \sqrt{n^2-1}/q$ \cite{Archambault:93}, where \emph{n} is the index of the glass and \emph{q} is the order of the resonance. Therefore, our fiber has a broad resonance-free zone from 250 nm to 450 nm in the UV. It is also of further advantage that the thin core-walls cause our fiber to operate in the second anti-resonance window ($1 < 2t \sqrt{n^2-1}/\lambda < 2$), since this is known to lead to higher coupling efficiency \cite{Fokoua:21,Zuba:22}. Achieving such thin core-walls is still a significant fabrication challenge.
\noindent In this article we report the fabrication and characterization of a hollow-core fiber (HCF) with a core-wall thickness of 220 nm for a targeted optimal guidance wavelength range of 300 nm to 320 nm. We evaluate the output beam quality by measuring the beam propagation ratio and investigate how bending the fiber impacts its transmission. Using the cutback-technique, we measured transmission loss of 0.13 dB/m at 320 nm and in addition we do not find an indication of fiber degradation when exposing the fiber to powers of up to 60 mW of UV light over tens of hours. The low measured loss is state of the art for the given spectral range. 

\section{Fiber fabrication}

\noindent The fiber was fabricated in a stack-and-draw process (for an overview, see \cite{Russell:06}). By adjustment of process parameters such as air-pressure applied inside the capillaries, furnace temperature and draw velocity, properties such as core diameter and wall thickness could be precisely controlled. We carefully chose optimal materials for the fiber production. In a previous study \cite{Rikimi:20}, unwanted residues of ammonium chloride were found on surfaces of fibers made from glass containing a high concentration of chlorine. These residues drastically reduce the fiber coupling efficiency. For the cladding capillaries we used Suprasil F310 with $<$ 0.2 ppm chlorine and $\sim$ 200 ppm hydroxyl content to prevent crystal growth on the surface of the fiber ends. High concentration of hydroxyl groups in the glass is known to improve the long-term transmission of light in the UV and VIS spectral range. The tube to which the capillaries were directly fused was made of HSQ100 fused silica with less than 20 ppm hydroxyl content. The outermost part of the fiber was Suprasil F300 which features a low amount of hydroxyl groups ($<$ 0.2 ppm) but a high amount of
chlorine (typically between 800 and 2000 ppm). Since all glass structures except for the capillaries are far away from the relevant core structure, the high amount of chlorine seems to not play any role and we did not observe unwanted residues. A strong suppression of higher-order modes (HOMs) to ensure single-mode transmission while simultaneously having relatively low bend losses was our primary target in designing the fiber. While optimal HOM-suppression requires
a ratio between capillary diameter \emph{d} and core diameter \emph{D} of \emph{d/D} = 0.68 \cite{Uebel:16}, the critical bend radius scales as $R_{cr} \sim \frac{\left(d/D\right)^2}{1-d/D}$ \cite{Frosz:17}, therefore we aimed for a \emph{d/D} value somewhat smaller than 0.68. For the presented fiber, the core diameter \emph{D} is about 18 µm and the capillary diameter \emph{d} is $\sim$ 9 µm, resulting in a \emph{d/D}-ratio of 0.5. The core-wall thickness of the capillaries is around 220 nm. Therefore, the first and second-order resonances can be expected at $\sim$ 450 nm and $\sim$ 250 nm with a broad transmission band in-between suitable for many applications in quantum labs.

\section{Experimental results}

\noindent We used two different laser sources to carry out the experiments. The measurement described in Sec. \ref{Exp:Cutback-method} was performed with a Diode Pumped Solid-State laser (DPSS, model: 4136) from the company LASOS. It emits 10 mW of CW power at a wavelength of 320 nm and the beam profile has a Gaussian shape with M\textsuperscript{2} = 1.04 $\pm$ 0.03. For the measurements in Secs. \ref{Exp:BendingProperties} and \ref{Exp:LifetimeInvestigation}, a frequency doubled dye laser system (Matisse 2DX, Sirah Lasertechnik GmbH) was used. 
We operated this setup at a wavelength of 302 nm and an optical power between 30 and 60 mW for degradation characterization of the fiber. The beam profile was elliptical with a ratio of 1:2. In both setups, no further spatial cleaning or filtering of the beams was done prior to fiber-coupling. To match the predicted mode diameter of the fiber, the waist of the collimated beam was enlarged by a factor of 7 (DPSS laser) or 10 (dye laser) to about 1.9 mm and then focused down to a beam diameter of $2\tilde{w_0}$ = 12 µm using a lens with a focal length of 125 mm where $\tilde{w_0}$ is the beam waist at the fiber facet. It is known that the coupling efficiency is maximized at all guided wavelengths when the beam diameter for fiber coupling is approximately 70\% of the core diameter \cite{Fokoua:21}. We used two flat mirrors to align the beam and optimize the transmission through the fiber (see Fig. \ref{Fig1:REM}(b)).

\subsection{Output beam quality}
\label{Exp:OutputBeamQuality}
\begin{figure}[t!]
\centering\includegraphics[width=2.5in]{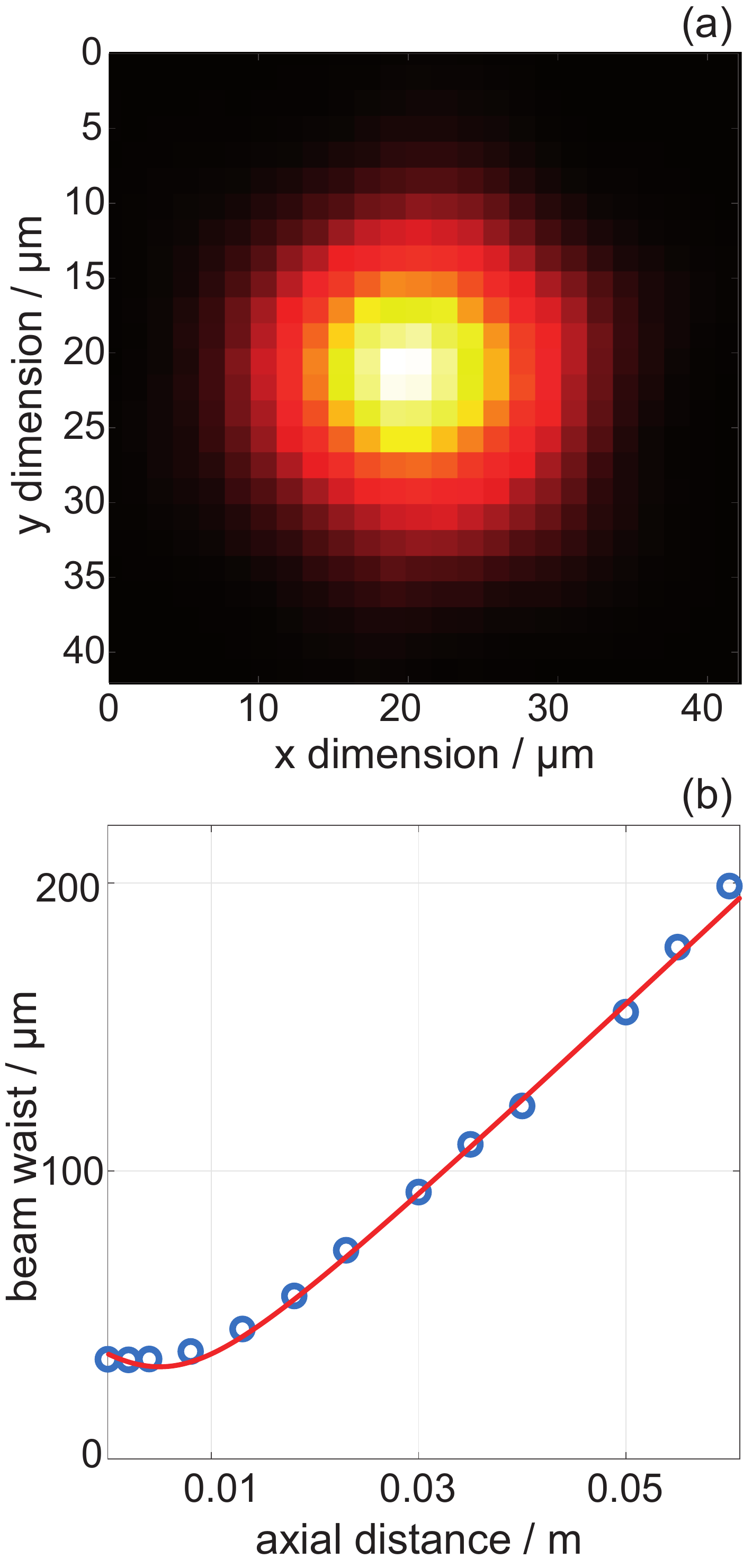}
\caption{(a) Image of the near-field intensity profile of the output beam. It shows a nearly symmetric beam shape. (b) In order to characterize the transverse mode properties of the output beam we perform beam waist measurements on the beam which has been focused to a waist of 30 µm. The plot shows the beam waist as a function of axial distance to the position of the focusing lens (blue dots). The red curve is a fit to the data and is used to extract the beam propagation ratio M\textsuperscript{2} = 1.03 $\pm$ 0.04.}
\label{Fig2:Image}
\end{figure}

\noindent We confirmed single-mode guidance by monitoring intensity profiles of the output beam with a CCD camera. In previous work \cite{Yu:18}, a HCF with $d/D \sim$ 0.43 showed a somewhat asymmetric mode profile and higher transmission loss at a wavelength of 355 nm. Figure \ref{Fig2:Image} (a) shows the near-field beam profile for the tested fiber. For this we used the DPSS system, but we were able to achieve equal results with the dye laser. The beam has a highly symmetric profile. The ratio of the diameters between the two main axes (y/x) is 0.98 which indicates a high quality circular shape. To further characterize the quality of the output beam, we measured the beam propagation ratio M\textsuperscript{2}. For this, we focused the beam with a lens and measured the radius of the beam at various axial distances near the focal point, see Fig. \ref{Fig2:Image}(b). Then, we used Gaussian beam fitting according to
\begin{equation}
w\left(z\right) = w_0\sqrt{\left(M^2+M^2\left(\frac{\lambda}{\pi w_0^2}\right)^2\left(z-z_0\right)^2\right)}
\end{equation} with $w$ as the measured radius at a position $z$ and the wavelength $\lambda$ = 320 nm. The beam waist $w_0$ and  a position offset $z_0$ are free fitting parameters. We extracted M\textsuperscript{2} = 1.03 $\pm$ 0.04. Note that for a single-mode $\mathrm{TEM_{00}}$ (Gaussian) laser beam, M\textsuperscript{2} = 1. Higher-order modes are suppressed by large propagation losses, so that the mode is cleaned by the  fiber after approximately 1 m. We noticed that fixation of the fiber end plays an essential role for optimal fiber coupling. In order to avoid mechanical stress on the fiber, we used a low expansion glue for fixation of the fiber on a V-groove holder. We could not find differences regarding the output beam quality between the two used laser sources within our measurement uncertainty. This additionally confirms the mode cleaning behavior of our HCF.

\subsection{Cutback-method}
\label{Exp:Cutback-method}

\begin{figure}[t!]
\centering\includegraphics[width=2.6in]{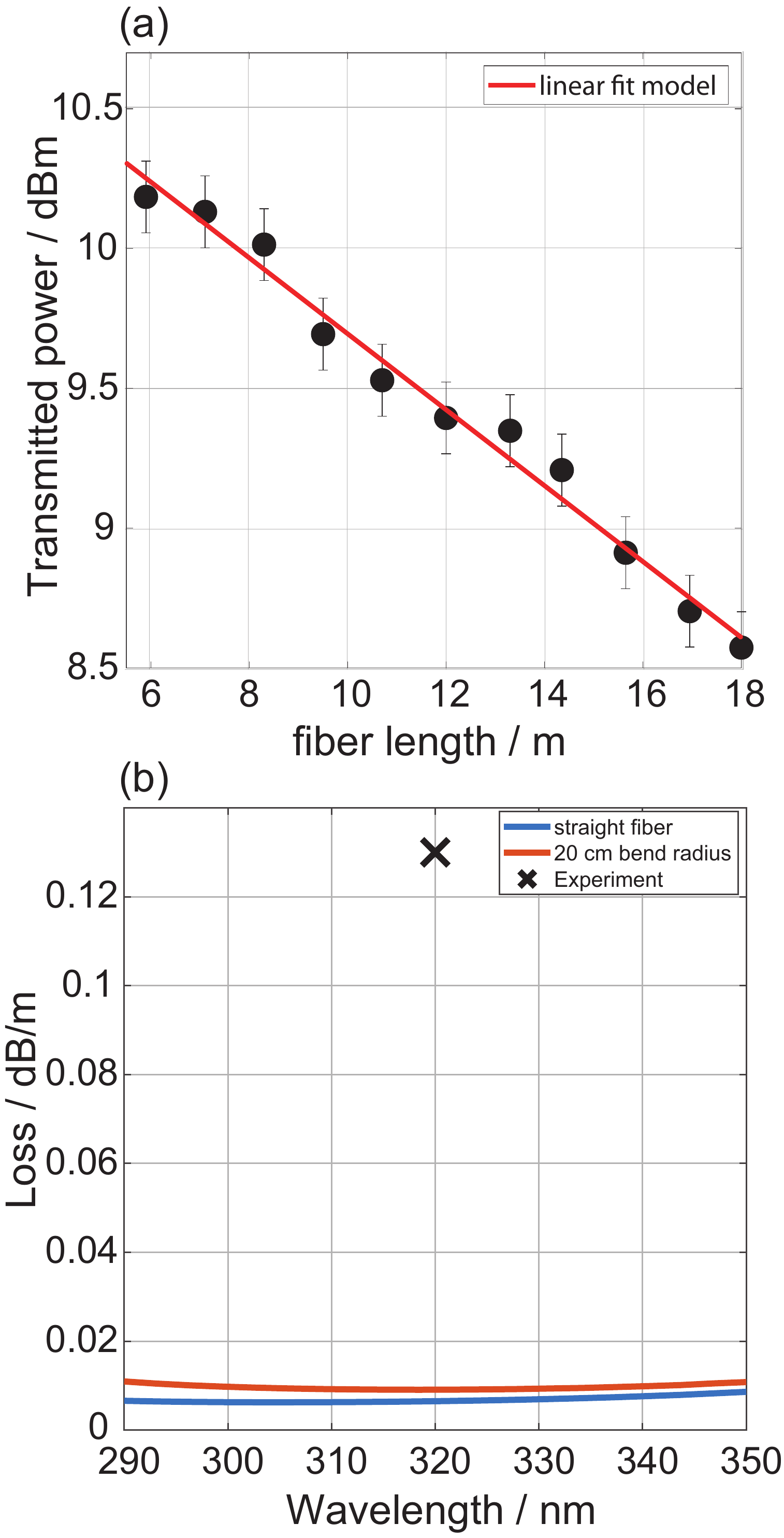}
\caption{(a) Cutback loss measurement at a
wavelength of 320 nm. Shown is the measured transmission power as a function of the fiber length. The error bars of the data points result from the uncertainty of the power measurement by a silicon photo detector. From a linear fit (red line), a transmission loss of 0.13 dB/m
is obtained and an in-coupling loss of only 6 \%. (b) Calculation of the transmission loss for a straight fiber (blue) and a fiber with a bend radius of 20 cm (red) which is typical for our experiments. The experimentally observed loss (black cross) is a factor of $\sim$ 10 larger than the calculation.}
\label{simulation}
\end{figure}

\noindent The propagation loss was measured using the multiple-cutback method without changing the coupling into the fiber. We kept the fiber straight for 30 cm in order to minimize bend-loss and wound the remaining length of the fiber on a coil with 20 cm radius. Once the transmission was maximized by adjusting the fiber coupling optics, the alignment was fixed. The results of the cutback-measurement are shown in Fig. \ref{simulation}. We measured a transmission power loss of 0.13 dB/m with an in-coupling loss of about 6\% extracted from a linear fit. We note that a 94\% coupling efficiency was also reported in Ref. \cite{Zuba:22}. Our results for transmission loss for a wavelength around 320 nm are very similar to the findings of \cite{Gao:18} and are significantly better than \cite{Yu:18}($\sim$ 0.6 dB/m) and \cite{Hartung:15}($\sim$ 3 dB/m). Still, the measured loss at 320 nm is found to be about an order of magnitude higher than predicted by finite element modelling (FEM), which considers only confinement and bend loss, see Fig. \ref{simulation}(b). The additional loss observed in the measurement is most likely due to surface scattering and microbending. Microbending is an effect which occurs due to small perturbations along the fiber whereas surface scattering arises from a thermodynamically unavoidable surface roughness \cite{Osorio:23}. Both effects increase significantly towards shorter wavelengths \cite{NumkamFokoua:23, Osorio:23} (surface scattering scales $\propto \lambda^{-3}$ and microbending scales $\propto \lambda^{-6}$). Within the fabrication process azimuthal asymmetries in the cladding capillaries occur along the fibre and FEM simulations show a strong sensitivity of loss resonances to the core-wall thickness.

\subsection{Performance under bending}
\label{Exp:BendingProperties}

\begin{figure}[t!]
\centering
\includegraphics[width=2.5in]{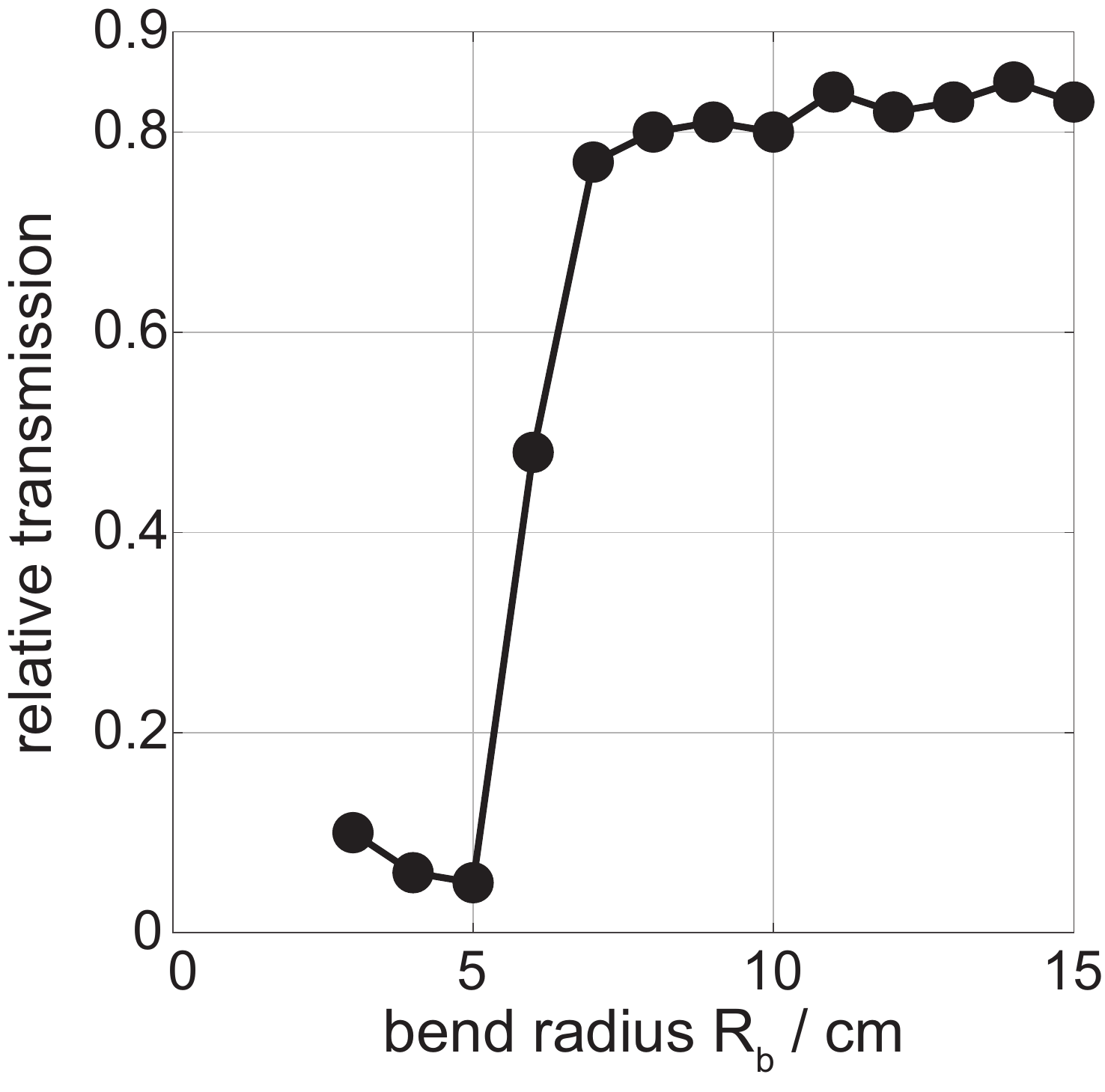}
\caption{Relative transmission as a function of the bend radius $R_{b}$ of a single winding of the fiber on a coil. The calculated critical bend radius $R_{cr}$ is approximately around 5.5 cm. The transmission drops abruptly below a bend radius of about 6 cm.}
\label{Exp:BendRadius}
\end{figure}

\noindent For practical applications, an important requirement for the fabricated fibers is to be bending insensitive at bend radii relevant for laboratory experiments. Below a certain critical bend radius, single-ring fibers and similar fiber types exhibit a strong drop in transmission due to bend-loss \cite{Frosz:17} and also the shape of the outgoing mode changes. At the critical radius, the fundamental core mode couples to the fundamental mode inside one of the cladding tubes, thereby experiencing high loss. We give an upper estimate for the critical bend radius using the simplified expression from \cite{Frosz:17}
\begin{equation}
    \frac{R_{cr}}{D} = \frac{D^2}{\lambda^2} \cdot \frac{\pi^2}{u_{01}^2}\frac{\left(d/D\right)^2}{1-d/D},
\end{equation}
where the value $u_{01}$ depends on the specific core mode. For the fundamental mode $u_{01} = 2.405$. In the original expression, there is an angular dependence which is neglected here. In practical lab situations it is not desirable to have to control the twist of the fiber carefully.
Furthermore we see that bend-sensitivity can be reduced by choosing suitably small values of \emph{D} and \emph{d/D}. However, \emph{d/D} should also not be chosen much smaller than \emph{d/D} = 0.68 in order not to lose some suppression of higher-order modes \cite{Uebel:16}. For our design parameters \emph{d/D} = 0.5, the critical bend radius is calculated to be $R_{cr} \approx$ 5.5 cm at $\lambda$ = 300 nm. The bending properties were measured with the remaining 5 m long piece of the same sample used for the cutback-method. For this, the initial part of the fiber is kept straight for about 30 cm to not change the fiber coupling efficiency by later adjustments of the bend radius. In the further course of the fiber, this is followed by a single loop the radius of which is varied (see Fig.\ref{Exp:BendRadius}). At a bend radius of about 6 cm, the transmission decreases abruptly and also the beam profile changes to a multi-mode pattern. This agrees quite well with the expectation from theory. 

\subsection{Lifetime investigation}
\label{Exp:LifetimeInvestigation}

\begin{figure}[t!]
\centering
\includegraphics[width=2.5in]{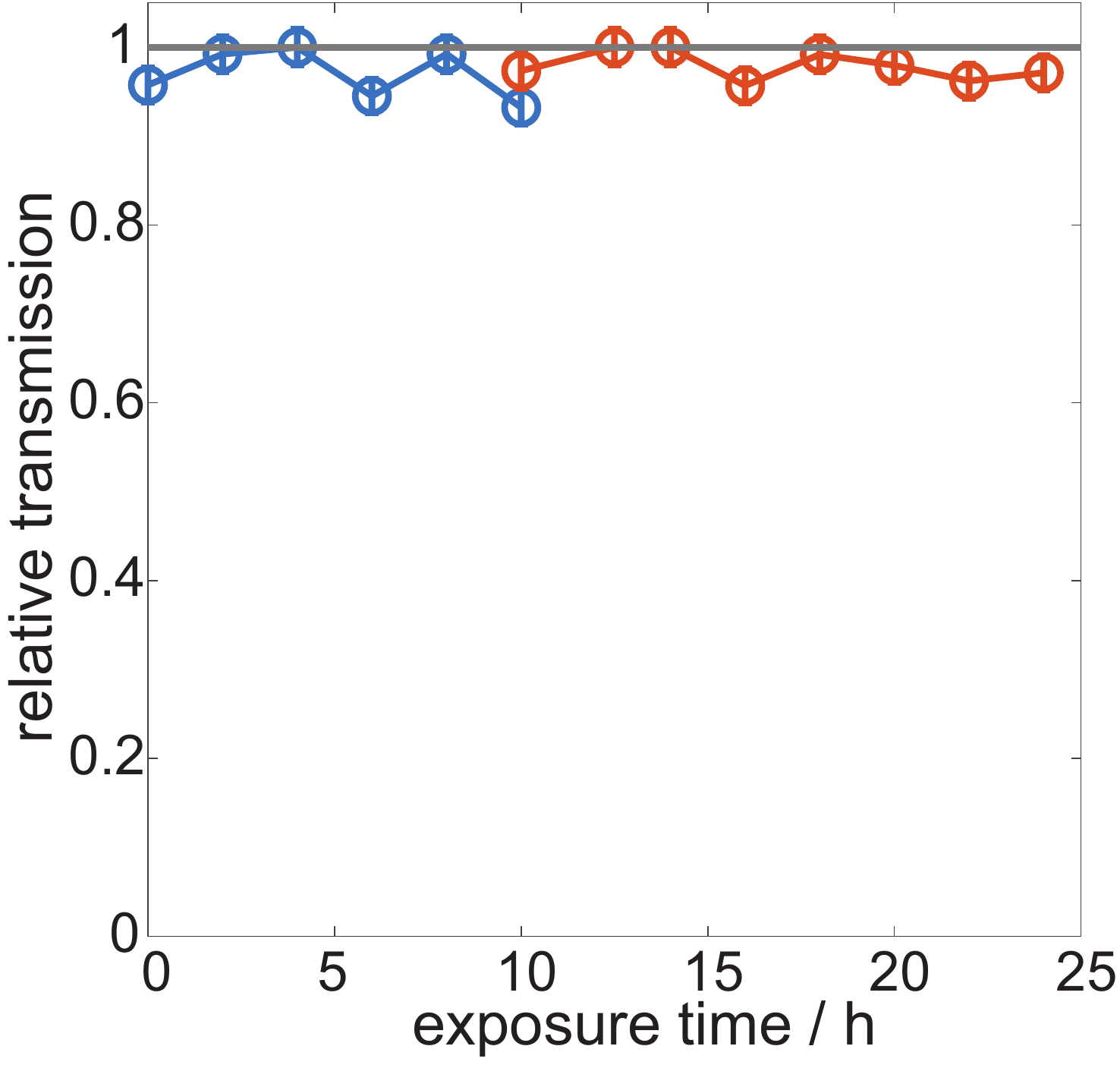}
\caption{Longterm exposure of the fiber at 302 nm with the dye laser system. We show relative transmissions over several hours of exposure for two different power levels. We normalized the transmission to the maximal observed transmission for each power level.
Blue (red) data points correspond to 30 (60) mW of laser power prior to the fiber coupling, respectively. Due to imperfections in the transverse mode, coupling was limited to about 65\%.}
\label{Exp:LongtermMeasurement}
\end{figure}

\noindent When exposed to UV light solid-core fibers exhibit a degradation of their transmission properties, a process which is known as solarization. Often this degradation already occurs within a few hours of exposure. We tested for solarization using laser powers of up to 60 mW. Figure \ref{Exp:LongtermMeasurement} shows the transmission of UV laser light during the course of one day. No degradation of the fiber transmission is observable. We also performed a 10 mW UV-light exposure of the fiber over several weeks, the data of which are not shown here. Again, no degradation was observable.

\section{Summary}

\noindent We characterized the transmission of UV laser light through a single-ring hollow-core optical fiber which is designed for low-loss, single-mode transmission over a wavelength range of 250 nm to 450 nm. Using wavelengths of 302 nm and 320 nm we found single-mode transmission with a mode quality of M\textsuperscript{2} = 1.03 and a loss of 0.13 dB/m using the cutback method. Our designed fiber has a transmission attenuation similar to other fiber types recently reported for UV guidance \cite{Gao:18} but operates in a much broader spectral range of 250 to 450 nm due to very thin core-walls of 220 nm. Furthermore, it operates in the second anti-resonant window leading to high fiber coupling efficiency. As the fiber has a critical bend radius of about 6 cm, it can be conveniently used in compact laboratory environments that are typical for atom or ion traps. The fiber shows no effects of degradation over tens of hours of effective exposure time with a power up to 60 mW. In principle, it should be possible to fabricate similar fibers but with additional polarization-maintaining properties (see \cite{Roth:18} for circular polarization for 1593 nm laser light). Such fibers are of high interest for applications e.g. in quantum optics labs where polarization stability is crucial.

\section{Acknowledgements}
\noindent This research was funded by the German Research Foundation (DFG) within the priority
program “Giant Interactions in Rydberg Systems” (DFG SPP 1929 GiRyd).
\bibliography{sample}
\balance
\end{document}